# A Technique to Measure Focal Length of a Lens with no Bulk Motion using Tunable Optics and Optical MEMS Technology

Syed Azer Reza[1] and Arslan Anjum[1]
1. Department of Electrical Engineering, Lahore University of Management Sciences, 54792, Lahore, Pakistan

**Abstract**

This paper presents a motion-free technique for characterizing the focal length of any spherical convex or concave lens. The measurement system uses a Gaussian Beam from a Laser Source (LS), an Electronically Controlled Variable Focus Lens (ECVFL), a Digital Micromirror Device (DMD) and a Photo-Detector (PD). As the proposed method does not involve any motion-stages or other moving components, the focal length is measured without requiring any mechanical motion of bulk components. The method requires measuring the spot size of the Gaussian Beam at the DMD plane for various settings of the ECVFL focal length. These beam spot size measurements, are used to estimate the focal length of a lens sample by employing standard polynomial-fitting techniques. Due to the inherent motion-free nature of the proposed setup, the measurements are fast, repeatable, reliable and ideal for use in industrial lens production, manufacturing of imaging systems and sensitive laboratory experiments. Using a DMD and an ECVFL also allows for automating the lens characterization process. As a proof-of-concept, experiments were performed to determine focal length values of multiple lens samples with the proposed technique. The values of focal lengths of all samples are in agreement with the focal length values which were provided by the manufacturer of these test lenses. The estimated focal length values are within the 1% manufacturing tolerance values that were provided.

# 1. Introduction

An accurate measurement of the focal length of a lens is critical in various scientific applications in optics. It is imperative to characterize lenses to a high degree of accuracy for several reasons. Precision optical imaging systems often employ multiple lenses and a precise knowledge of the focal length of each individual lens is vital to achieve the desired accuracy and performance. A reliable knowledge of the focal length value and the associated manufacturing tolerance of custom-made lenses is also critical as end-users rely on the accuracy of these values when designing high-performance optical systems. Furthermore, for large-scale industrial production of lenses, it is highly desirable that the technique for characterizing lens samples is simple and fast. The simplicity and speed of the lens characterization technique is desirable in order to cope with potentially high production rates of lens samples. In addition, it is also highly desirable that the method to determine focal length can be automated. Process automation ensures repeatability of measurements and further helps in enhancing the rate at which lenses can be characterized.

Generally, the focal length of a lens is estimated using techniques that rely on recording multiple measurements. Techniques that involve multiple measurements minimize errors by mitigating the effect of highly inaccurate independent data points through the estimation of a mean and standard deviation. The process of averaging is widely known to be equivalent of a smoothing operation and therefore the contribution of severely inaccurate data points is minimized. Moreover, measurement errors can also be statistically estimated for any multiple-point dataset instead of a single measurement value. Hence, it is highly desirable to estimate the focal length of any lens through the use of a multiple-measurement technique because calculating a mean value of multiple measurements eliminates errors which are mostly associated with single data point measurement techniques.

Typically, test-benches rely on the motion of various optical elements to make multiple measurements in order to estimate the focal length of a lens sample. These motion-based techniques require precision motion control of bulk optical components, thus compromising system robustness and speed. Motion-based techniques also experience mechanical hysteresis of stages and movable mounts which might result in performance degradation with use over time. Thus it is highly desirable that bulk-motion of mechanical components is minimized or, even better, eliminated in a focal length measurement test-bench.

Various methods for focal length characterization have been proposed in prior art. Techniques using Talbot Interferometry generally rely on measuring Moiré Fringe tilt angles and displacements to estimate the focal length [1-4]. Similarly, it was shown that the focal length can be estimated by measuring the Lao Phase interferometric fringe intensity distribution [5-6]. These techniques are motion-based that use translation stages to move gratings and mirrors along the optical axis to obtain multiple measurements. Moreover, determining changes in fringe patterns requires high-fidelity image processing techniques with a large post-processing overhead which results in an increased image post-acquisition processing time of characterizing a single lens sample.

Other reported interferometry-based techniques include focal length characterization using Dammann Gratings [7], Fizeau Interferometry [8-9], Shearing Interferometry [10], and Fresnel

Interference Holograms [11]. These techniques also require mechanical motion of bulk components for multiple measurements. Some recently proposed focal length characterization techniques include interferometric as well as non-interferometric systems. These methods include focal length measurement through phase-shifted Moiré deflectometry [12], double grating interferometry [13-14] and observation of spot patterns and spherical aberrations [15]. All these techniques offer varying degrees of accuracy but nonetheless, require bulk motion of mechanical components. Recently, motion-free methods using tunable optics have been proposed to determine the propagation characteristics of Gaussian Beams [16-17] and the central location of the Gaussian intensity distribution [18]. The method that we present in this paper achieves motion-free characterization of the focal length of any spherical lens with the use of an electronically controlled tunable lens and a MEMS-based DMD.

The proposed measurement test-bench does not involve bulk mechanical motion and instead a fixed optical setup records multiple measurements and determines the focal length of a lens sample. The measurement technique solely relies on the micro-motion of liquid inside the ECVFL and the micro-mirrors of a DMD. A mean value of focal length and the associated measurement errors are estimated through systematic measurements of beam spot size at the DMD plane for various settings of the ECVFL focal length. In the following sections, the design of the motion-free system is discussed and a validation of concept is presented through experimental results. The process was entirely automated which shows that this method of measuring focal length is ideal for deployment in industrial lens production units.

## 2. Theory of the proposed measurement technique

The method that we propose, characterizes the focal length $f_{Stat}$ of a spherical lens $L_1$ as shown in Fig.1. An optical beam with wavelength $\lambda$ exits the LS and passes through an ECVFL which is placed at a fixed distance of $D_4$ from the exit aperture of LS. A point 'P', located at a distance $D_5$ behind the exit aperture of LS, defines the location of the $1/e^2$ minimum beam waist $w_0$ inside the laser cavity. The total distance $D_1$ from point 'P' to the ECVFL plane is therefore given by $D_1 = D_4 + D_5$. The test lens $L_1$ is placed between the ECVFL and the DMD. The distance of the test lens $L_1$ is fixed at $D_2$ from the ECVFL and $D_3$ from the DMD. The beam after passing through the ECVFL and $L_1$ is normally incident at the plane of a reflective DMD. The lenses $L_1$ and the ECVFL are carefully aligned such that the beam passes through the center of both the lenses.

The DMD is a reflective spatial light modulator formed from a two-dimensional array of digitally-controlled two-state micro-mirrors. Each micro-mirrors is individually controllable and can be set to either an angle of $+\theta$ degrees or $-\theta$ degrees. A Photo-Detector (PD) is positioned such that it collects the fraction of optical power of the Gaussian beam which is incident on the DMD micro-mirrors which are set to the $+\theta$ state. The use of the DMD with a PD in such an arrangement, allows digital characterization of the spatial profile of a Gaussian beam at the DMD plane through a moving knife-edge pattern on the DMD [19]. A DMD-based beam profiler is a motion-free design which can be used to digitally profile high irradiance Gaussian Beams involving only the micro-motion of the DMD micro-mirrors.

The positions of LS, ECVFL, $L_1$ and the DMD are fixed and remain unaltered during the measurements. Therefore the distances $D_2$ and $D_3$ are fixed and known. $D_1$ is also fixed depending on the beam propagation properties of the LS. Even if $D_1$ is unknown, it can be determined using the same set of data which is used to calculate $f_{Stat}$. The measurement of focal length does not rely on any prior knowledge of the properties of the LS i.e., location of point 'P' (and hence $D_1$) and the minimum beam waist size $w_0$ at 'P'.

In order to determine $f_{Stat}$, the ECVFL focal length $f$ is tuned to different values within the ECVFL operating range. The RMS amplitude of the voltage signal from the ECVFL controller to the ECVFL controls and sets its focal length. The $1/e^2$ spot size $w_{out}$ on the DMD plane is measured for different settings of $f$ and these values of $w_{out}(f)$, are used to estimate $f_{Stat}$. If the laser beam profile of LS is Gaussian, then the cross-sectional electric field intensity of the beam at a distance 'z', along the optical axis, from point 'P' is given by [20]:

$$E(r,z) \propto \exp\left(jkr^2 \big/ 2q(z)\right) \qquad (1)$$

Where q(z) is the complex q-parameter of the Gaussian beam and it is given by:

$$\frac{1}{q(z)} = \frac{1}{R(z)} - j\frac{\lambda}{\pi w^2(z)} \qquad (2)$$

Here $R(z)$ and $w(z)$ are the radius of curvature and the $1/e^2$ beam waist radius respectively at a distance z along the optical axis from the minimum beam waist point 'P', $k = 2\pi/\lambda$, and $r = \sqrt{x^2 + y^2}$, where $(x,y)$ are the Cartesian coordinates of the optical field plane. At point 'P', $R = \infty$, hence the complex q-parameter $q(z)$ is purely imaginary and it is equal to $q_{in}$ and expressed as:

$$\frac{1}{q_{in}} = -j\frac{\lambda}{\pi w_0^2} \Rightarrow q_{in} = j\frac{\pi w_0^2}{\lambda} = jZ_R \qquad (3)$$

Here $Z_R$ is the Rayleigh range of the LS. The q-parameter at the DMD plane is $q_{out}$ and it is expressed as:

$$\frac{1}{q_{out}} = \frac{(Cq_{in} + D)}{(Aq_{in} + B)} = \frac{(jCZ_R + D)}{(jAZ_R + B)} \qquad (4)$$

The imaginary part of $1/q_{out}$ is obtained by separating the real and imaginary parts in Eq.4. This leads to:

$$\operatorname{Im}\left\{\frac{1}{q_{out}}\right\} = -\frac{Z_R(AD - BC)}{A^2 Z_R^2 + B^2} \qquad (5)$$

We also know from Eq.2 that

$$\operatorname{Im}\left\{\frac{1}{q_{out}}\right\} = -\frac{\lambda}{\pi w_{out}^2} \qquad (6)$$

Therefore we conclude that

$$\frac{\lambda}{\pi w_{out}^2} = \frac{Z_R(AD - BC)}{A^2 Z_R^2 + B^2} \qquad (7)$$

and as $AD - BC = 1$ for ray trace matrices

$$w_{out}^2 = \frac{\lambda}{\pi} \frac{(A^2 Z_R^2 + B^2)}{Z_R} \qquad (8)$$

The ABCD Matrix of the proposed system is given as:

$$\begin{bmatrix} A & B \\ C & D \end{bmatrix} = \begin{bmatrix} 1 & D_3 \\ 0 & 1 \end{bmatrix} \times \begin{bmatrix} 1 & 0 \\ -1/f_s & 1 \end{bmatrix} \times \begin{bmatrix} 1 & D_2 \\ 0 & 1 \end{bmatrix} \times \begin{bmatrix} 1 & 0 \\ -1/f & 1 \end{bmatrix} \times \begin{bmatrix} 1 & D_1 \\ 0 & 1 \end{bmatrix} \qquad (9)$$

Here

$$A = \left(1 - \frac{D_3}{f}\right)\left(1 - \frac{D_2}{f_{Stat}}\right) - \frac{D_3}{f_{Stat}} \qquad (10)$$

$$B = D_1\left\{\left(1 - \frac{D_3}{f}\right)\left(1 - \frac{D_2}{f_{Stat}}\right) - \frac{D_3}{f_{Stat}}\right\} + D_2\left(1 - \frac{D_3}{f}\right) + D_3 \qquad (11)$$

$$C = \left(\frac{1}{f}\right)\left(\frac{D_2}{f_{Stat}} - 1\right) - \frac{1}{f_{Stat}} \tag{12}$$

$$D = \left(1 - \frac{D_2}{f}\right) - \left(\frac{D_1}{f}\right)\left(1 - \frac{D_2}{f_{Stat}}\right) - \left(\frac{D_1}{f_{Stat}}\right) \tag{13}$$

Substituting $A$, $B$, $C$ and $D$ in Eq.7, we derive the following relationship between the $1/e^2$ spot radius $w_{out}(f)$ at the DMD plane for an ECVFL focal length setting of $f$.

$$w_{out}^2(f) = w_0^2 \left\{ \left[1 - \frac{D_2 + D_3}{f} - \frac{D_3}{f_{Stat}} + \frac{D_2 D_3}{ff_{Stat}}\right]^2 + \left[\frac{\lambda}{\pi w_0^2}\left(D_1 + D_2 + D_3 - \frac{D_1(D_2 + D_3)}{f} - \frac{D_3(D_1 + D_2)}{f_{Stat}} + \frac{D_1 D_2 D_3}{ff_{Stat}}\right)\right]^2 \right\} \tag{14}$$

As is evident from Eq.14, if the values of $D_1$, $D_2$, $D_3$, $\lambda$ and $w_0$ are fixed and known, it is sufficient to calculate the value of $f_{Stat}$ if the value of $w_{out}(f)$ is accurately measured for any single setting of $f$. Although it is possible to determine $f_{Stat}$ using a single measurement of $w_{out}(f)$, making multiple measurements of $w_{out}(f)$ for various settings of $f$ enables us to estimate a mean value of $f_{Stat}$ and the measurement error in the least squares sense. Recording multiple measurements simply requires electronically changing the focal length of the ECVFL and measuring the corresponding beam spot radius $w_{out}(f)$. Also, the method does not depend on any prior knowledge of the properties of LS; $w_0$ and $D_1$. Through multiple measurements of $w_{out}(f)$ for ECVFL settings of $f$, and the use of Eq.14, a system of simultaneous equations is obtained through measured data. All unknown quantities of the system can be determined by solving this system of equations.

Another photo-detector $PD_2$ can be positioned to capture reflected power from the DMD micro-mirrors in the $-\theta$ state. Simultaneous power measurements with PD and $PD_2$ are only required if there are significant power fluctuations in the beam that is produced by LS [20].

## 3. Errors in the Estimation of Focal Length

The focal length $f_{Stat}$ of a lens sample is determined through the least-squares estimation using Eq.14. The focal length $f$ of the ECVFL is considered to be the independent variable while the beam radius $w_{out}(f)$ at the DMD plane is measured for each setting of the ECVFL focal length. Hence $w_{out}(f)$ is the dependent variable for our purpose of focal length estimation. The residual of the least-squares fit is the difference between the theoretical predicted value according to the model predicted by Eq.14 and the measured value of $w_{out}(f)$ for each setting of $f$. Therefore the residual for each data point ($w_{out}$ versus $f$) is given by:

$$(\Delta w_{out})_i = \left| (w_{out\_i})_{Theoretical} - (w_{out\_i})_{Measured} \right| \tag{15}$$

The best estimate for $f_{Stat}$ is $(f_{Stat})_{Estimated}$ and it is obtained when the sum of the squares of the residuals is minimized. We compare this best estimate value with the individually calculated values $(f_{Stat})_i$ for every measurement data point $\left( f_i, (w_{out\_i})_{Measured} \right)$. Here the subscript '$i$' denotes the $i^{th}$ measurement. From Eq.14, we know that

$$w_{out\_i}^2 = w_0^2 \left\{ \left[ 1 - \frac{D_2 + D_3}{f_i} - \frac{D_3}{(f_{Stat})_i} + \frac{D_2 D_3}{f_i (f_{Stat})_i} \right]^2 + \left[ \frac{\lambda}{\pi w_0^2} \left( D_1 + D_2 + D_3 - \frac{D_1(D_2 + D_3)}{f_i} - \frac{D_3(D_1 + D_2)}{(f_{Stat})_i} + \frac{D_1 D_2 D_3}{f_i (f_{Stat})_i} \right) \right]^2 \right\} \tag{16}$$

We approximate the error in our estimation $\langle \Delta f_{Stat} \rangle$ of $f_{Stat}$ by calculating the average difference between the individually estimated values $(f_{Stat})_i$ and the least-squares best estimate $(f_{Stat})_{Estimated}$. The average estimation error for $N$ data points is subsequently given by:

$$\langle \Delta f_{Stat} \rangle = \frac{\sum_{i=1}^{N} \left[ (f_{Stat})_{Estimated} - (f_{Stat})_i \right]}{N} \tag{17}$$

## 4. Experimental demonstration of Test bench

As a proof-of-concept, the focal length measurement test bench of Fig.1 was implemented in order to estimate the focal length of three lenses: two bi-convex lenses 'Lens 1' and 'Lens 2' with focal lengths $f_{Stat}$ = 100mm ±1% (Thorlabs LB1676) and $f_{Stat}$ = 75mm ±1% (Thorlabs LB1901) respectively and a bi-concave lens 'Lens 3' with $f_{Stat}$ = −100mm ±1% (Thorlabs LD1613$^c$). All values of the focal lengths of these lenses and their respective tolerances were provided by the lens manufacturer, Thorlabs Inc. The aim of our experiments was to accurately determine the central focal length values of these three lens samples to within the tolerance values that are provided by the lens manufacturer.

In the experimental setup the distances $D_2$ and $D_4$ were approximately set to 4.9cm and 41.0cm respectively and the distance $D_3$ was roughly set to 8.6cm. The laser properties i.e. values of $D_5$ and $w_0$ were determined during the lens characterization measurements instead of using known values. The proposed method works regardless, and previously known values of $w_0$ and $D_5$ can also be used instead. A red He-Ne laser was used for the experiment with $\lambda$ = 632.8nm and a half angle beam divergence $\theta_B$ = 1.7mrad. The estimated value of minimum waist radius was $w_0$ =117.8μm and its location behind laser aperture was known to be $D_5$ = 14.5cm. For $D_4$ = 41.0cm and a known value of $D_5$, $D_1$ = 55.5cm.

The DMD used for the measurements was part of an Acer C20 LED projector with a micro-mirror pitch of 7.56μm. The ECVFL used was the Optotune EL-10-30 lens. In figures 2, 3 and 4, the theoretical values of $w_{out}(f)$ for different ECVFL focal length settings of $f$ are plotted for each of the three lens samples respectively for 14cm > $f$ > 5.5cm focal length range. The theoretically estimated values of $w_{out}(f)$ for different settings of $f$ is plotted for Lens 1 with $f_{Stat}$ = 100mm ±1% in Fig.2, for Lens 2 with $f_{Stat}$ = 75mm ±1% in Fig.3 and Lens 3 with $f_{Stat}$ = -100mm ±1% in Fig.4.

Actual measurements of $w_{out}(f)$ were recorded for various settings of $f$ for all three lens samples. The measured values of $w_{out}(f)$ for different settings of $f$ are summarized for each lens sample in Table.1. The measurement data points for Lens 1, Lens 2 and Lens 3 are plotted in Fig.2, Fig.3 and Fig.4 respectively. The dots in Figs.2-4 represents actual data from the experiments. For each of the sample lenses, the data points lie in close proximity to the theoretical curves. Furthermore, the measurement data is used to estimate $f_{Stat}$. The fitted curve on the data set for each lens sample is obtained through standard curve-fitting techniques. We used the Levenberg–Marquardt Algorithm to obtain a least squares curve fit on the recorded data for each lens sample. The fitted curves are also plotted in Figs.2-4 for each of the three lens samples respectively. The estimated values of $f_{Stat}$ and percentage difference with respect to the manufacturer-provided central focal length values for the three lens samples is summarized in Table.2. For each of the three lens samples, it is evident that there is a strong resemblance between the fitted curves from experimental measurements, and the corresponding theoretical plots. This is proof of an excellent agreement between the theory presented and the actual experimental results.

As summarized in Table.2, the focal length of all lenses were characterized very accurately within the manufacturing tolerance limit of ±1% provided by Thorlabs Inc meaning thereby that the measured and the manufacturer-provided central focal length values differ by less than 1%. By using data in Table.1, the values of the focal length was experimentally estimated to be 99.1mm, 75.6mm and -99mm for Lens 1, Lens 2 and Lens 3 respectively. The difference between the central focal length value provided by the manufacturer and the experimentally measured focal length differ by 0.9%, 0.8% and 1% for the three lens samples.

The manufacturer provided for each lens sample a central focal length value with a ±1% manufacturing tolerance limit with respect to this central value, therefore it was not possible to determine an absolute measurement error in $f_{Stat}$ because calculating absolute error requires a single absolute focal length value which was not available for any of the three sample lenses. Hence it is better to calculate the estimation error that is always present in methods that involve an estimation-through-regression approach such as the method of least-squares.

In order to measure $w_{out}(f)$ at the DMD plane for any setting of $f$, the DMD and the PD were interfaced to a personal computer through an indigenous Labview application. This application implemented a moving knife-edge on the DMD and recorded optical powers at the PD for various positions knife edge positions. Using this data, the horizontal $w_H(f)$ and vertical $w_V(f)$ $1/e^2$ beam radii were determined and the values averaged to determine a mean $1/e^2$ beam radius $w_{out}(f)$ presented in Table.1. The process was repeated for various focal length settings of the ECVFL to obtain sufficient data points for a least-squares estimate of $f_{Stat}$. This method can also be used to characterize cylindrical lenses; in which case, $w_{out}(f)$ is only measured along the axis of optical focus. The general method of focal length characterization remains the same.

## 5. Estimation Error in Measured Focal Lengths of the Lens Samples

For the experiments performed on Lenses 1, 2 and 3 we calculate the estimation error for the focal length values that were determined using the method that we propose. Table.3 summarizes the errors in the focal length values due to limitations with the use of the least-squares curve-fitting technique. Using Equations (15)-(17), the estimation errors are calculated to be 0.38 mm, 0.0306 mm and -0.34 mm for Lens 1, Lens 2 and Lens 3 respectively. These values of the estimation errors correspond to percentage measurement errors of 0.39 %, 0.04 % and 0.35 % for the three lens samples used in our experiments.

## Conclusion

This paper presents a novel method of measuring the focal length of any lens using tunable optics and a MEMS-based DMD. The method uses the DMD to make measurements of the beam spot size for various settings of the ECVFL. A least-squares curve-fit operation is performed on the data to estimate the focal length of the sample. The method does not require any motion of bulk optical components. This motion-free design, when used with an automated digital data acquisition and post-processing system, enhances speed, efficiency, repeatability and reliability of the method. Experimental results are in excellent agreement with theory. Various lens samples were characterized to within the manufacturing tolerance limits. The proposed method can be potentially useful for various commercial and academic applications in industrial optical production and development of imaging systems.

# Figures

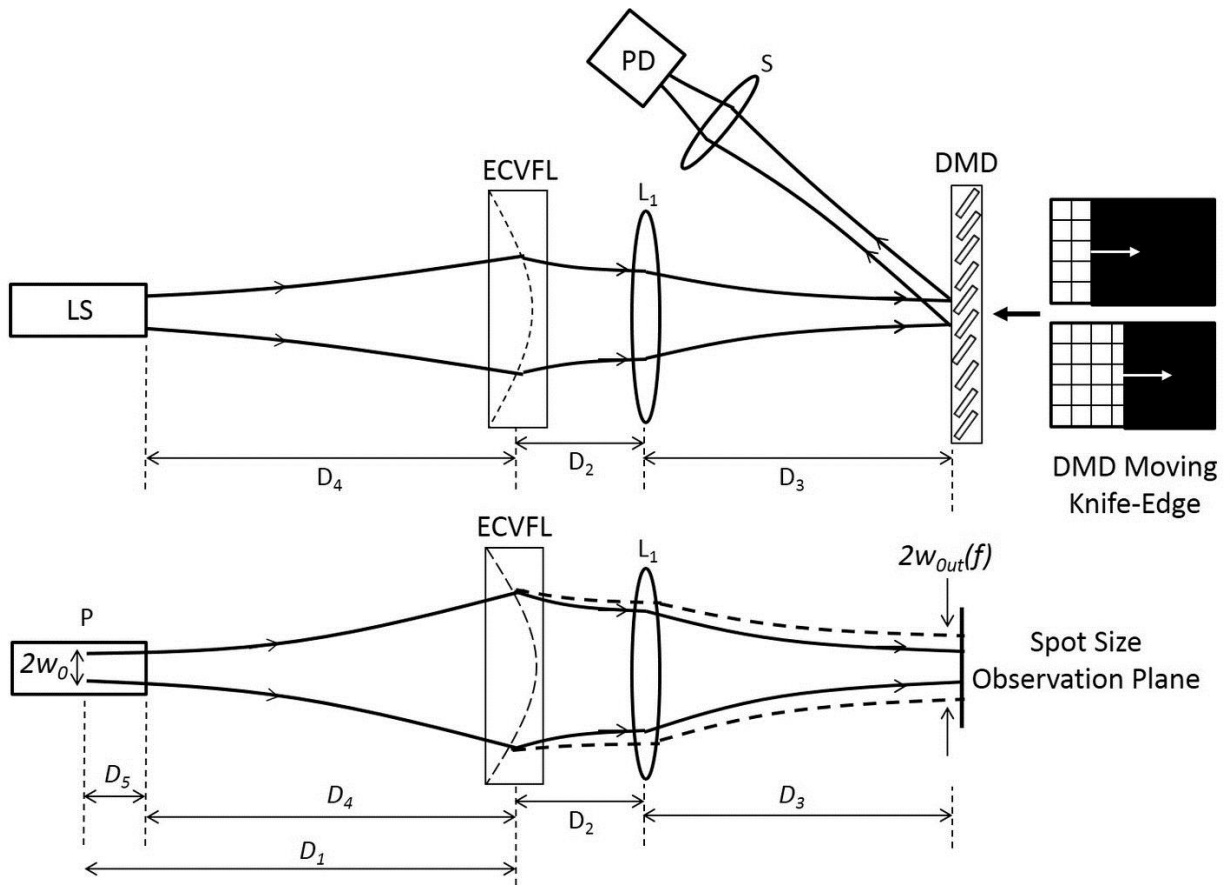

Fig.1 Design of the proposed motion-free optical work-bench for the characterization of the focal length $f_{Stat}$ of a static sample lens $L_1$. Setup includes an ECVFL, a DMD, a Photo-Detector (PD), and a Laser Source (LS).

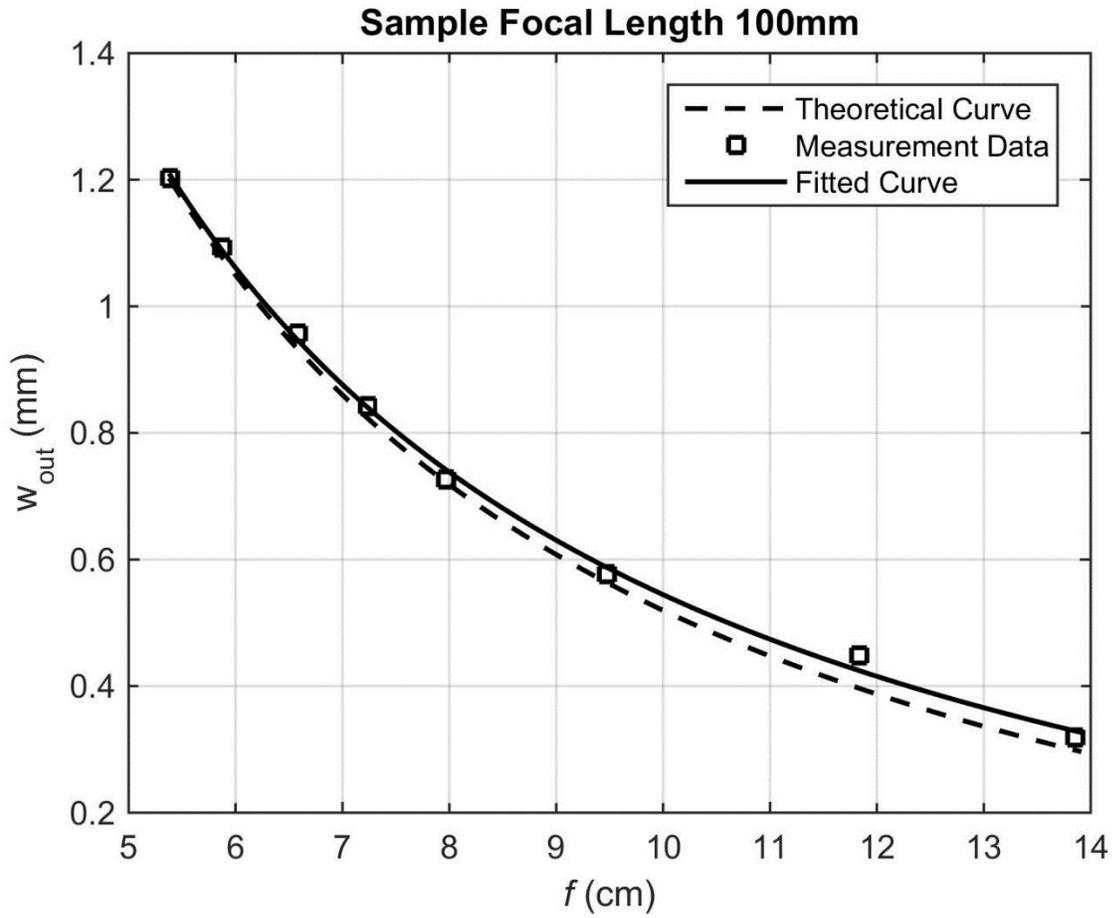

Fig.2 Theoretical variation of $w_{out}(f)$ with $f$ for Lens 1 with $f_{Stat}$ = 100mm and the comparison to experimental data points and the resulting fitted curve.

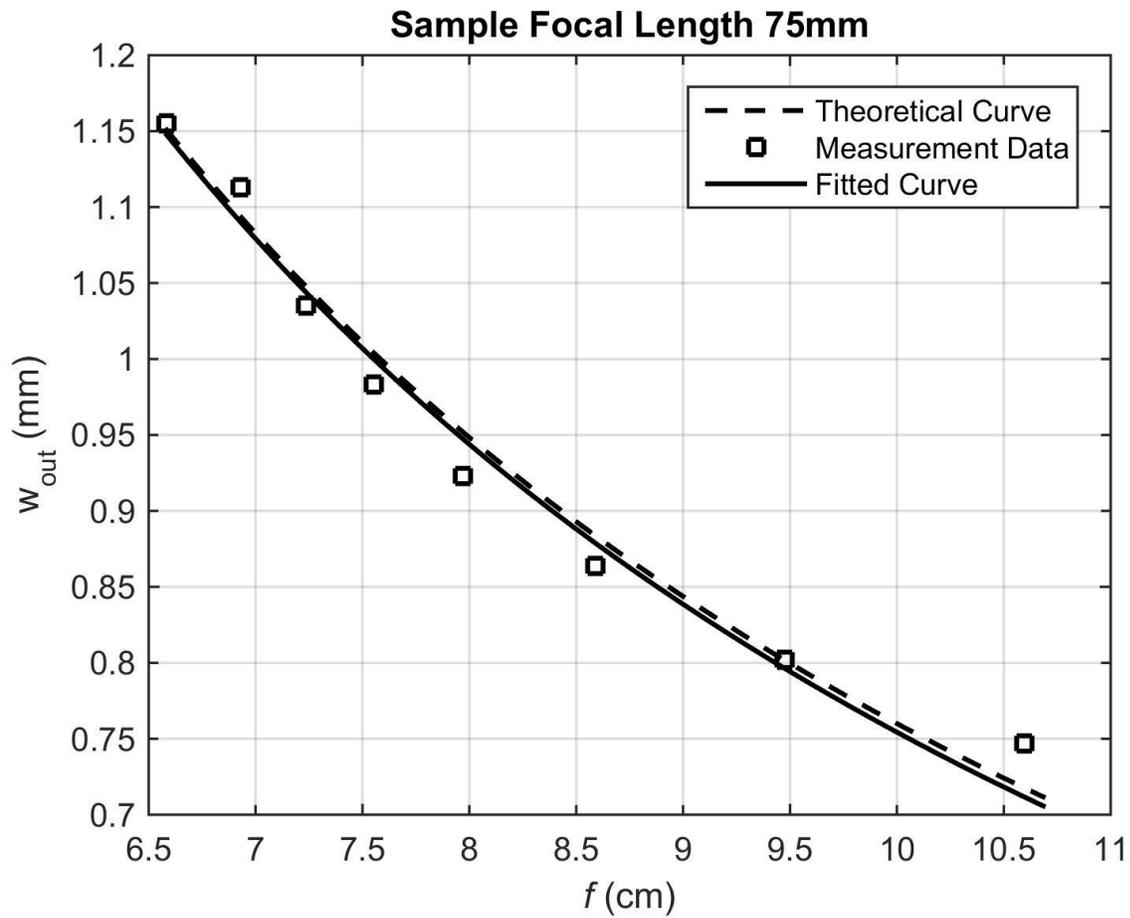

Fig.3 Theoretical variation of $w_{out}(f)$ with $f$ for Lens 2 with $f_{Stat}$ = 75mm and the comparison to experimental data points and the resulting fitted curve.

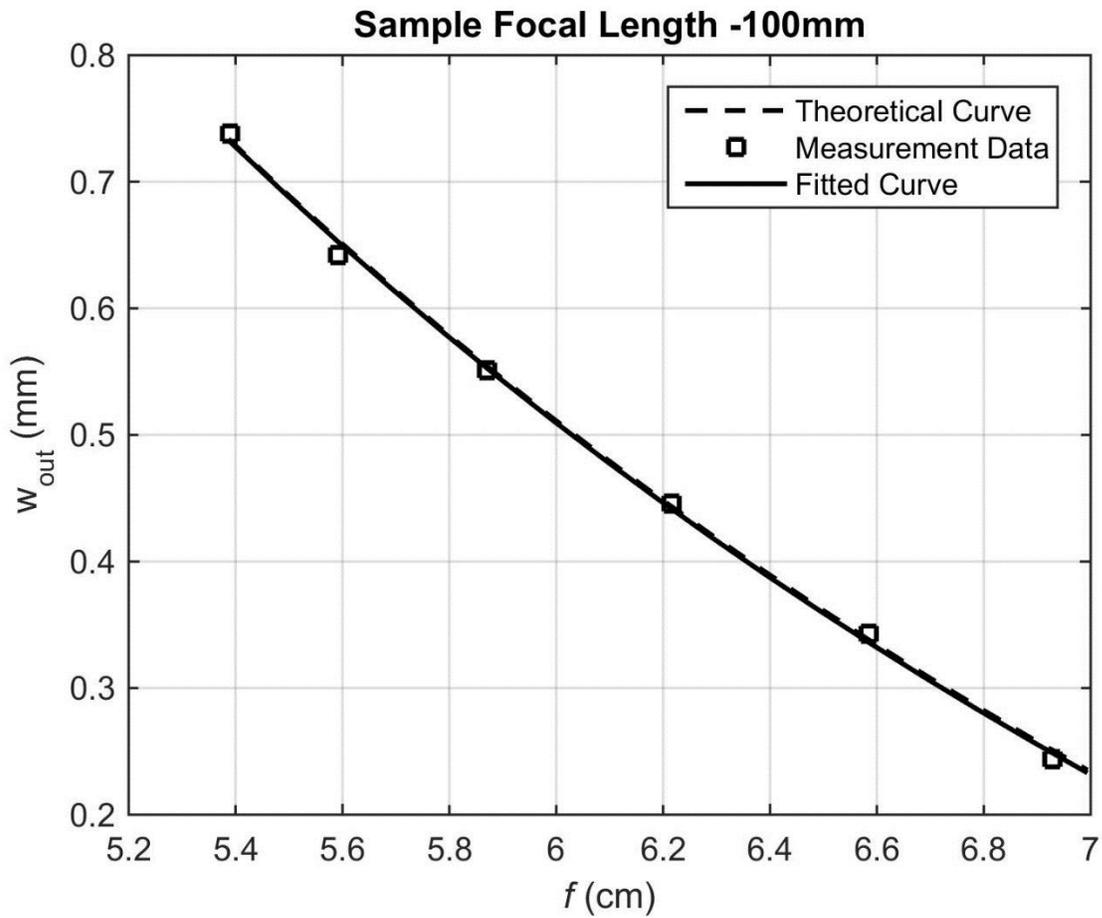

Fig.4 Theoretical variation of $w_{out}(f)$ with $f$ for Lens 3 with $f_{Stat}$ = -100mm and the comparison to experimental data points and the resulting fitted curve.

# Tables

| Lens 1 ($f_{Stat}$ = 100mm ± 1%) | | Lens 2 ($f_{Stat}$ = 75mm ± 1%) | | Lens 3 ($f_{Stat}$ = -100mm ±1%) | |
|---|---|---|---|---|---|
| ECVFL Focal Length $f$ (cm) | Measured $1/e^2$ Beam Radius $w_{out}(f)$ (mm) | ECVFL Focal Length $f$ (cm) | Measured $1/e^2$ Beam Radius $w_{out}(f)$ (mm) | ECVFL Focal Length $f$ (cm) | Measured $1/e^2$ Beam Radius $w_{out}(f)$ (mm) |
| 13.8562 | 0.319 | 10.5967 | 0.747 | 6.9286 | 0.244 |
| 11.8333 | 0.449 | 9.4743 | 0.802 | 6.5843 | 0.343 |
| 9.4743 | 0.577 | 8.5901 | 0.864 | 6.2163 | 0.446 |
| 7.9698 | 0.726 | 7.9698 | 0.923 | 5.8710 | 0.551 |
| 7.2369 | 0.842 | 7.5531 | 0.983 | 5.5919 | 0.642 |
| 6.5843 | 0.957 | 7.2369 | 1.035 | 5.3900 | 0.738 |
| 5.8710 | 1.093 | 6.9286 | 1.113 | | |
| 5.3900 | 1.202 | 6.5843 | 1.155 | | |

Table.1 Data of Measured Beam Radius $w_{out}(f)$ at different ECVFL Focal Length Settings for all three Lens Samples.

| Lens Number | Manufacturer Values $f_{Stat}$ (mm) ± % Tolerance | Measured $f_{Stat}$ (mm) | Difference between Measured and Manufacturer Provided Values of $f_{Stat}$ |
|---|---|---|---|
| 1 | 100 ±1% | 99.1 | 0.9 % |
| 2 | 75 ±1% | 75.6 | 0.8 % |
| 3 | -100 ±1% | -99.0 | 1.0 % |

Table 2. Comparison of Manufacturer-Provided and Measured Data for Lenses

| Estimation Property | Sample Lens 1 | Sample Lens 2 | Sample Lens 3 |
|---|---|---|---|
| Goodness of Fit ($R^2$) | 0.99887 | 0.98200 | 0.99808 |
| Average Residual | 1.34 x$10^{-08}$ $m^2$ | 3.13 x$10^{-08}$ $m^2$ | 5.98 x$10^{-09}$ $m^2$ |
| $f_{Stat}$ Estimation Error | 0.38 mm | 0.0306 mm | -0.34 mm |
| Percentage Estimation Error | 0.39 % | 0.04 % | 0.35 % |

Table.3 The Goodness of Fit and Least-Squares Estimation Error in the Measured Focal Length Values $f_{Stat}$ for all three Lens Samples.

# Figure Captions

Fig.1 Design of the proposed motion-free optical work-bench for the characterization of the focal length $f_{Stat}$ of a static sample lens $L_1$. Setup includes an ECVFL, a DMD, a Photo-Detector PD, and a Laser Source LS.

Fig.2 Theoretical variation of $w_{out}(f)$ with $f$ for Lens 1 with $f_{Stat}$ = 100mm and the comparison to experimental data points and the resulting fitted curve.

Fig.3 Theoretical variation of $w_{out}(f)$ with $f$ for Lens 2 with $f_{Stat}$ = 75mm and the comparison to experimental data points and the resulting fitted curve.

Fig.4 Theoretical variation of $w_{out}(f)$ with $f$ for Lens 3 with $f_{Stat}$ = -100mm and the comparison to experimental data points and the resulting fitted curve.

# Table Captions

Table.1 Data of Measured Beam Radius $w_{out}(f)$ at different ECVFL Focal Length Settings for all three Lens Samples.

Table 2. Comparison of Manufacturer-Provided and Measured Data for Lenses

Table.3 The Goodness of Fit and Least-Squares Estimation Error in the Measured Focal Length Values $f_{Stat}$ for all three Lens Samples.